\documentclass[preprint,showpacs,preprintnumbers,amsmath,amssymb]{revtex4}

\usepackage{graphicx}
\usepackage{dcolumn}
\usepackage{bm}
\usepackage{epsfig}

\begin{document}

\title{ Unified description of pairing, trionic and quarteting states
  for one-dimensional $SU(4)$  attractive fermions}

\author{ X.W. Guan$^{\dagger }$, M.T. Batchelor$^{\dagger  \ddagger}$,
 C. Lee$^{\star}$ and J.Y. Lee$^{\dagger }$}
\affiliation{${\dagger}$ Department of Theoretical Physics, Research
 School of Physics and Engineering,
Australian National University, Canberra ACT 0200,  Australia}
\affiliation{${\ddagger}$ Mathematical Sciences Institute,
Australian National University, Canberra ACT 0200,  Australia}
\affiliation{${\star}$ Nonlinear Physics Centre and ARC Centre of
Excellence for Quantum-Atom Optics, Research School of Physics 
and Engineering,Australian National University, Canberra ACT
 0200, Australia }

\date{\today}

\begin{abstract}
Paired states, trions and quarteting states  in one-dimensional $SU(4)$ attractive 
fermions are investigated via exact Bethe ansatz calculations. 
In particular, quantum phase transitions are identified and calculated  from the
quarteting phase into normal Fermi liquid, trionic states and spin-2 paired states 
which belong to the universality class of linear field-dependent magnetization in the 
vicinity of critical points.
Moreover, unified exact results for the ground state
energy, chemical potentials and complete phase diagrams for isospin $S=1/2, \,1, \,3/2$
attractive fermions with external fields are presented.
Also identified are the magnetization plateaux of $m^z=M_s/3$ and $m^z=2M_s/3$, where 
$M_s$ is the magnetization saturation value.
The universality of finite-size corrections and collective
dispersion relations provides a further test ground for low energy effective
field theory.
\end{abstract}



\maketitle

Experimental advances with higher spin fermionic systems of ultracold
atoms present a unique opportunity to rigorously test the current
understanding of molecular superfluids and more generally to probe the
nature of quantum many-body systems.
Three-component ultracold fermions give rise to a phase transition
from three-body bound states called trions into the BCS pairing 
state \cite{Rapp,Lecheminant,Demler}.
In particular, spin-3/2 interacting atomic fermions have been predicted
to exhibit a quarteting phase, i.e., a bound state of two BCS pairs
\cite{Zhang,Lecheminant,Tsvelik}, and BCS pairing with total spin-$2$ \cite{Ho}.
Such molecular superfluids have currently received considerable
interest in the context of the one-dimensional (1D) lattice Hubbard model
of ultracold atoms \cite{Lecheminant,Lecheminant2,Hofstetter} and  degenerate
quantum gases \cite{GBLZ,Hu,Ortiz}  due to new progress
in experimental realization of highly degenerate atomic gases \cite{Li}.

Furthermore, recent experiments on systems of
ultracold atoms confined to one dimension (1D) \cite{1D-F,weiss,g2} have 
revived interest in Bethe ansatz (BA) integrable models of interacting bosons and
multi-component fermions.
The nature of ``no diffraction'' in the many-body scattering matrix of 1D
integrable models results in key features of quantum many-body
physics which are specified by the phenomena of spin-charge
separation, scaling dimensions and universality classes of quantum
phase transitions and criticality \cite{Takahashi-B,Tsvelik-book,Giamarchi-b}.
In this Letter, we study complete phase diagrams and quantum phase
transitions in integrable 1D $SU(4)$ attractive fermions with external magnetic fields.
We demonstrate that quantum phase transitions from quarteting states
into phases of normal Fermi liquid, trions and spin paired states are
fully controlled by Zeeman splittings.
In particular, the unified exact results obtained for isospin $S=1/2,\,1,\,3/2$
attractive fermions display a simplicity and universality which 
gives insight in understanding high spin paired states and spin liquid behaviour 
in multi-component interacting fermions.

{\it The model.} We consider a $\delta$-function (contact potential) interacting system of
$N$ atomic fermions with equal mass $m$ which may occupy four
possible hyperfine levels ($|i\rangle$, $i=1,\cdots,4$) labeled by 
isospin states $\pm 3/2,\,\pm1/2$ and constrained by
periodic boundary conditions to a line of length $L$.
The Hamiltonian \cite{Sutherland,Takahashi} reads
${\cal{H}}={\cal{H}}_0+{\cal{H}}_I+E_z$ with kinetic energy
${\cal{H}}_0=-\frac{\hbar ^2}{2m}\sum_{i = 1}^{N}\frac{\partial
^2}{\partial x_i^2}$, interaction energy ${\cal{H}}_I=g_{\rm 1D}
\sum_{1\leq i<j\leq N} \delta (x_i-x_j)$ and Zeeman energy
$E_z=\sum_{i=1}^4N^{i}\epsilon^{i}_Z(\mu_B^{i},B)$.
Here $N^{i}$ is the number of fermions in state $| i\rangle$
with Zeeman energy $\epsilon^{i}_Z$ (acting as  a chemical potential
$\mu_i$ \cite{Demler}) determined by the magnetic
moments $\mu_B^{i}$ and the magnetic field $B$.
The spin-independent contact interaction $g_{\rm 1D}$ remains between
fermions with different hyperfine states and preserves the spins in
each hyperfine states, i.e., $N^i$ with $i=1,\ldots,4$ are good quantum
numbers.
Although these conditions appear somewhat restrictive, utilizing the
broad Feshbach resonances, it is possible to tune scattering
lengths between different states close to each other to form high
degeneracy Fermi gases \cite{Grimm,Li,1D-F}.
Thus the model still captures the essential physics relevant to the 
multiple phases of molecular superfluids. 
The coupling constant $g_{\rm 1D} =-{\hbar ^2 c}/{m}$ with
interaction strength $c=-{2}/{a_{\rm 1D}}$ determined by the
effective 1D scattering length $a_{\rm 1D}$ \cite{Olshanii}.
For simplicity, we choose the dimensionless units of $\hbar = 2m = 1$ 
and use the dimensionless coupling constant $\gamma=c/n$ with linear
density $n ={N}/{L}$.

\begin{figure}
{{\includegraphics
[width=0.23\linewidth,angle=-90]{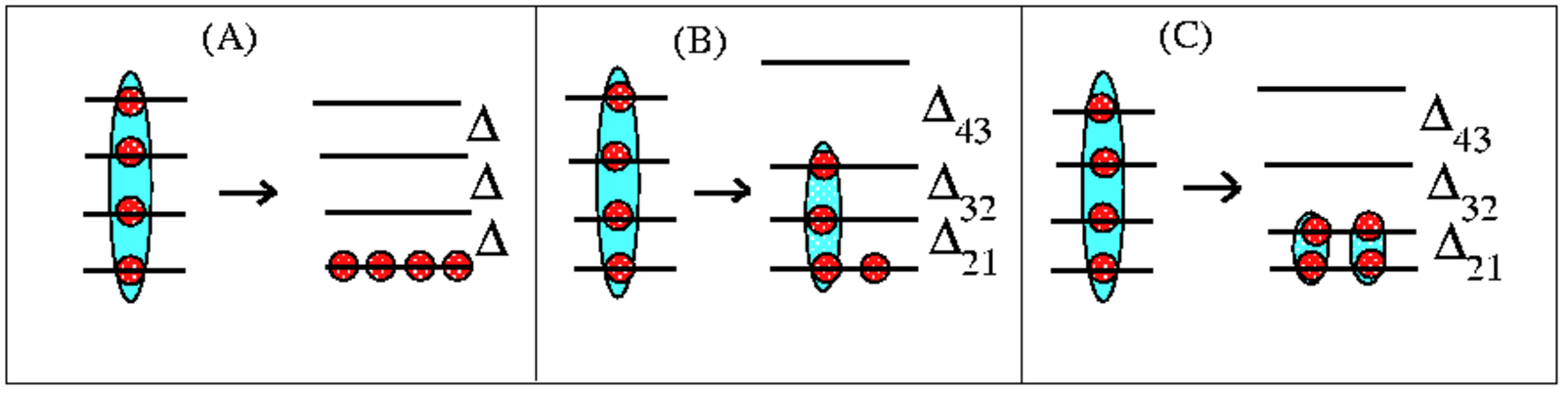}}}
\caption{Depiction of phase transitions from quarteting states into
  (A) normal Fermi liquid, (B) trions and (C) bound pairs.  Ellipses
  denote spin-neutral bound states.}\label{fig:states}
\end{figure}

For an irreducible representation $\left[4^{N_4}3^{N_3}2^{N_2}1^{N_1} \right]$, a 
four-column Young tableau encodes the numbers of unpaired
fermions ($N_1$), bound pairs ($N_2$), trions ($N_3$) and quarteting
states ($N_4$) with $N_i=N^i-N^{i+1} $ and $N^5=0$.  
For convenience in calculation, we rewrite the Zeeman energy as
$E_Z=-\sum_{i=0}^3H_iN_i$.  Here $H_0=\sum_{i=1}^4\epsilon_z^i/4$ is
irrelevant because $N_0=N$ is fixed. 
The other values $H_1,H_2,H_3$ are related
to Zeeman splittings $\Delta_{i+1\,i}=\epsilon^{i+1}_Z-\epsilon^{i}_Z$
with $i=1,2,3$ via the relation
\begin{equation}
\left(\begin{array}{l}
H_1\\
H_2\\
H_3
\end{array}
\right)=\frac{1}{4}\left(\begin{array}{lll}
3&2&1\\
2&4&2\\
1&2&3
\end{array}\right)\left(\begin{array}{l}
\Delta_{21}\\
\Delta_{32}\\
\Delta_{43}
\end{array}
\right).\label{H-Z}
\end{equation}
We shall find that equally spaced (linear) Zeeman splitting,
i.e., $\Delta_{a+1\,a}=\Delta$ for $a=1,2,3$, drive a smooth phase
transition from a quarteting phase into a normal Fermi liquid, as 
depicted in part (A) in Fig~ \ref{fig:states}.  
Unequally spaced (nonlinear)
Zeeman splittings may trigger spin-neutral bound states which are
illustrated in parts (B) and (C) in Fig~ \ref{fig:states}.
It follows from the relation (\ref{H-Z}) that for linear Zeeman splitting, the
magnetic moments of a trion, a bound pair and the unpaired fermion are
$\frac{3}{2}$, $2$ and $\frac{3}{2}$, respectively. The quarteting
state remains a spin singlet.

The energy eigenspectrum is given in terms of the quasimomenta
$k_j$ of the fermions via $E=\sum_{j=1}^Nk_j^2$, which
satisfy the BA equations \cite{Sutherland,Takahashi}
\begin{eqnarray}
\exp(\mathrm{i}k_{j}L)&=&\prod_{l=1}^{M_{1}}\frac{k_{j}-\Lambda^{(1)}_{l}
+\mathrm{i} \frac12 c}{k_{j}-\Lambda^{(1)}_{l}-\mathrm{i} \frac12 c}\nonumber\\
\prod_{\beta=1}^{M_{\ell-1}}\frac{\Lambda_{\alpha}^{(\ell)}-\Lambda_{\beta}^{(\ell-1)}
+\mathrm{i} \frac12 c}{\Lambda_{\alpha}^{(\ell)}-\Lambda_{\beta}^{(\ell-1)}-\mathrm{i} \frac12 c}
&=&-\prod_{\gamma=1}^{M_{\ell}}\frac{\Lambda_{\alpha}^{(\ell)}-\Lambda_{\gamma}^{(\ell)}+\mathrm{i}c}
{\Lambda_{\alpha}^{(\ell)}-\Lambda_{\gamma}^{(\ell)}-\mathrm{i}c}
\prod_{\delta=1}^{M_{\ell+1}}\frac{\Lambda_{\alpha}^{(\ell)}-\Lambda_{\delta}^{(\ell+1)}-\mathrm{i} \frac12 c}
{\Lambda_{\alpha}^{(\ell)}-\Lambda_{\delta}^{(\ell+1)}+\mathrm{i} \frac12 c}. \label{BE}
\end{eqnarray}
Here $j=1,\ldots, N$ and $\alpha=1,\ldots,M_{\ell}$.
The parameters $\Lambda_{\alpha}^{(\ell)}$ with
$\ell=1,2,3$ are the spin rapidities, where we denote 
$\Lambda_{\alpha}^{(0)}=k_{\alpha}$ and $\Lambda_{\alpha}^{(5)}=0$.
The quantum numbers  are given by $M_i=\sum_{j=i}^{3}N_{i+1}$, $M_0=N$.

{\it Charge bound states.} For attractive interaction, the BA equations 
(\ref{BE}) admit charge bound states and spin strings. 
In particular, the $SU(4)$ symmetry acquires three kinds of charge bound states:
quarteting states, trions and bound pairs.
The patterns of these bound states and spin strings determined by (\ref{BE}) 
underpin the nature of quantum statistics and many-body effects in the atomic system.

In the weak coupling regime, i.e., $L|c|\ll 1$, we find 
that  the imaginary parts $\mathrm{i}y$ of the charge bound
states are the roots of Hermite polynomials $H_k$ of degree $k$.
Specifically, $H_k(\frac{L}{2|c|}y)=0$, with $k=2,3,4$ for a bound pair, a
trion and a quarteting state, respectively. 
This result is indicative of a universal signature of many-body cooperative effects driven
by dynamical interaction. 
The real parts of the quasimomenta deviate slightly from the values
determined by Fermi statistics for the $c=0$ case.
In this weak coupling limit, the BA equations (\ref{BE}) reduce to Gaudin model-like
BA equations \cite{Gaudin} which describe the multiple charge bound state scattering.
Using these root patterns, we explicitly obtain the energy
(in units of $\hbar^2/2m$)
\begin{eqnarray}
&&\frac{E}{L} \approx \frac{\pi^2}{3}\sum_{k=1}^{\kappa}kn_k^3+\pi^2
\sum_{j=1}^{\kappa-1}jn_j\sum_{k=j+1}^{\kappa}n_k\sum_{\ell=j}^{\kappa}n_{\ell} \nonumber\\
&& \qquad
-|c|\left(2\sum_{i=1}^{\kappa}\sum_{j=i+1}^{\kappa}i(j-1)n_in_j+\sum_{j=1}^{\kappa}j(j-1)n_j^2\right).
\label{E-w}
\end{eqnarray}
This result unifies the ground state energy results for two-, three- and
four-component fermions (where $\kappa=2,3,4$, respectively) for weakly 
attractive interaction.
In this result the densities of unpaired fermions, BCS pairs, trions, and
qarteting states are denoted by  $n_a=N_a/L$ with $a=1,2,3,4$,
respectively.
The ground state energy (\ref{E-w}) is dominated by the kinetic energy
of composite particles and unpaired fermions and has a mean field
theory configuration, where the interaction energy accounts for
density density interaction between  charge bound states and between
charge bound states and unpaired fermions.  
For the weak coupling limit, spin-neutral bound states are unstable
against thermal and spin fluctuations.
They form a gapless superconducting phase.

On the other hand, for the strongly attractive regime $L|c|\gg 1$, 
the imaginary parts of the bound states become equal-spaced, i.e., 
a qarteting state has the
form $k_i=\Lambda^{(3)}_i\pm \mathrm{i} 3|c|/2$, 
$\Lambda^{(3)}\pm \mathrm{i} |c|/2 $; 
the trion state is $k_j=\Lambda_j^{(2)}\pm \mathrm{i}|c|, \lambda_j^{(2)}$ 
and for the bound pair $k_{r}=\Lambda_{r}^{(1)}\pm \mathrm{i}|c|/2$. 
The corresponding binding energies are given by 
$\epsilon_{\ell} = {c^2\ell(\ell^2-1)}/{12}$ with $\ell=1,2,3$, respectively.
Substituting these bound state root patterns into the BA equations (\ref{BE}), 
we explicitly obtain their real parts $\Lambda^{\ell}_{i,j,r}$, 
with $\ell=1,2,3$. 
This leads to a unified expresion 
\begin{eqnarray}
& &\frac{E}{L} \approx
  \sum_{k=1}^4\frac{\pi^2n_k^3}{3k}\left(1+\frac{2}{|c|}A_k+\frac{3}{c^2}A_k^2\right)-
  \sum_{\ell=2}^4n_{\ell} \epsilon_{\ell}\label{E}
\end{eqnarray}
for the ground state 
energy for two-, three- and four-component attractive fermions (in units of $\hbar^2/2m$)
where the functions $A_1=4n_2+2n_3+\frac{4n_4}{3}$,
$A_2=2n_1+n_2+\frac{8n_3}{3}+\frac{3n_4}{2}$,
$A_3=\frac{2n_1}{3}+\frac{16n_2}{9}+n_3+\frac{92n_4}{45}$ and
$A_4=\frac{n_1}{3}+\frac{3n_2}{4}+\frac{23n_3}{15}+\frac{11n_4}{12}$.
The functions $A_k$ reveal the scattering signature
in different channels, for example, $n_1$ does not appear in $A_1$ due to the 
lack of s-wave scattering for unpaired fermions.
We note that for two-component attractive fermions the terms involving 
$n_3$ and $n_4$ should be excluded \cite{Wadati} whereas for three-component
attractive fermions $n_4$ does not appear.
The unified structure of the ground state energy (\ref{E}) 
can be amended with appropriate $A_k$ functions for higher spin fermions.

{\it Charge bound states in equilibrium.}
The BA equations (\ref{BE}) in principle give the complete quantum states of the model.
However, at finite temperatures, the true physical states become degenerate.
In the thermodynamic limit, $L,N \to \infty$ with $N/L$ fixed,
the grand partition function is given by
$Z=\mathrm{tr}(\mathrm{e}^{-\cal{H}/T})=\mathrm{e}^{-G/T}$
\cite{Y-Y,Takahashi-B,Schlot}, where the Gibbs free energy $G=E +E_{\rm Z}-\mu
N-TS$,  the chemical potential $\mu$, the Zeeman energy
$E_{\rm Z}=-\sum_{i=0}^3H_iN_i$ and the entropy $S$ are given in terms
of densities of charge bound states and spin strings subject to the BA  equations (\ref{BE}).
The equilibrium
states are determined by the minimization of the Gibbs free energy,
which gives rise to a set of coupled nonlinear integral equations ---
the thermodynamic Bethe ansatz (TBA) equations.
Following the TBA treatment for spin-1/2 attractive fermions
\cite{Takahashi-B}, we obtain the dressed energy equations
\begin{eqnarray}
\epsilon^{(4)}(\tau)&=&4(\tau^2-\mu^{\rm 4})-a_3*{\epsilon^{(1)}}(\tau) -\left[a_{2,4}\right]*{\epsilon^{ (2)}}(\tau)\nonumber\\
& &-\left[a_{1,3,5}\right]*{\epsilon^{ (3)}}(\tau)-\left[a_{2,4,6}\right]*{\epsilon^{ (4)}}(\tau),\nonumber\\
\epsilon^{(3)}(\lambda)&=&3(\lambda^2-\mu^{\rm 3})-a_2*{\epsilon^{(1)}}(\lambda) -\left[a_{1,3}\right]*{\epsilon^{ (2)}}(\lambda)\nonumber\\
& &-\left[a_{2,4}\right]*{\epsilon^{ (3)}}(\lambda)-\left[a_{1,3,5}\right]*{\epsilon^{ (4)}}(\lambda),\nonumber \\
\epsilon^{(2)}(\Lambda)&=&2(\Lambda^2-\mu^{\rm 2})-a_1*{\epsilon^{
    (1)}}(\Lambda)-a_2*{\epsilon^{2}}(\Lambda) \nonumber\\
& &-\left[a_{1,3}\right]*{\epsilon^{(3)}}(\Lambda)-\left[a_{2,4}\right]*{\epsilon^{(4)}}(\Lambda),\nonumber\\
\epsilon^{ (1)}(k)&=&k^2-\mu^{\rm 1}-\sum_{i=1}^3a_i*{\epsilon^{ (i+1)}}(k),
\label{TBA-F}
\end{eqnarray}
which  will be  used to study  quantum phase
transitions at zero temperature.
In these equations the function
$a_j(x)=\frac{1}{2\pi}\frac{j|c|}{(jc/2)^2+x^2}$ and
$a_j*{\epsilon^{(a)}}(x)=\int_{Q_a}^{Q_a}a_j(x-y){\epsilon^{(a)}}(y)dy$
is the convolution. 
Furthermore, we have used the abbreviation $a_{i,j,k}=a_i+a_j+a_k$.
We denote the dressed energies $\epsilon^{( a)}$
with $ a = 1,\ldots,4$ and the effective chemical potentials $\mu^{
\rm a }=H_a/a+\mu+c^2(a^2-1)/(12)$ with $H_4=0$ for unpaired fermions,
pairs, trions and quarteting states, respectively.  
The negative part of the dressed energies $\epsilon^{(a)}(x)$ for
$x\le Q_{a}$ corresponds to the occupied states in the Fermi seas with the
positive part of $\epsilon^{(a)}$ corresponding to the unoccupied states.
The integration boundaries $Q_{m}$ characterize the ``Fermi
surfaces'' at $\epsilon^{(m)}(Q_{m})=0$.

The Gibbs free energy
per unit length at zero temperature is given by 
$G =\sum_{m=1}^4\frac{m}{2\pi}\int_{-Q_m}^{Q_m}{\epsilon^{(m)}}(x)dx$.
The dressed energy energy equations (\ref{TBA-F}) describe the band
fillings with respect to Zeeman splittings $H_i$ and chemical
potentials and provide complete phase diagrams and quantum phase transitions
for the model. 
Solving the equations (\ref{TBA-F})
by iteration among the relations 
$-\frac{\partial G}{\partial \mu} =n\,,\,-\frac{\partial G}{\partial H_i}=n_i,\,\,i=1,2,3$
and the Fermi point conditions $\epsilon^{\rm (m)}(Q_{m})=0$
gives the effective chemical potentials
\begin{eqnarray}
&&\frac{\mu^{\rm
    \kappa}}{\pi^2}\approx\frac{n_{\kappa}^2}{k^2}\left(1+\frac{2A_{\kappa}}{|c|}+\frac{3A_{\kappa
    }^2}{c^2}\right)+\frac{\vec{B}_{\rm
    \kappa} \cdot \vec{I}}{|c|}+\frac{3\vec{B}_{\kappa} \cdot
    \vec{A}}{c^2}\label{mu}
\end{eqnarray}
which characterize Fermi surfaces of stable spin-neutral states and
unpaired states.
In this equation we have introduced an inner product 
$\vec{B}_k\cdot \vec{A}$ with $\vec{A}=(A_1,A_2,A_3,A_4)$ 
and 
$\vec{B}_{\kappa}=(B_{\kappa}^1,B_{\kappa}^2, B_{\kappa}^3,B_{\kappa}^4)$.
Here 
$\vec{B}_{ 4}=
(\frac{2n_1^3}{9},\frac{n_2^3}{8},\frac{46n_3^3}{405},\frac{11n_4^3}{288})$,
$ \vec{B}_{ 3}=
(\frac{4n_1^3}{9},\frac{8n_2^3}{27},\frac{2n_3^3}{27},\frac{23n_4^3}{270})$,
$ \vec{B}_{ 2}=
(\frac{4n_1^3}{3},\frac{n_2^3}{6},\frac{16n_3^3}{81},\frac{n_4^3}{16})$ 
and 
$ \vec{B}_{1}= (0,\frac{2n_2^3}{3},\frac{4n_3^3}{27},\frac{n_4^3}{18})$. 
The $A_i$ are as given above and $\vec{I}$ is the identity vector. 
The expression (\ref{mu}) for the chemical potentials $\mu^{\rm k}$ contains the 
previous results for isospin $S=1/2,\,1$ Fermi gases \cite{GBLB,Wadati,GBLZ}.
We have found that the energy (\ref{E}) derived from the discrete BA equations for arbitrary
population imbalances can also be obtained from $E/L=\mu n+G +n_1 H_1+n_2 H_2+n_3H_3$.
This indicates that the BA spin-neutral states comprise the
equilibrium stable states in the thermodynamic limit.

{\it Magnetism and quantum phase transitions.}
The low-energy excitations split into collective excitations carrying
charge and collective excitations carrying spin. 
This leads to the
phenomenon of spin-charge separation. 
The charge excitations are
described by sound modes with a linear dispersion.  
The spin excitations are gapped \cite{Lecheminant2,Tsvelik} with a
dispersion relation $\epsilon_{\nu}(p)=\sqrt{\Delta_{\nu}^2+v_{\nu}^2p^2}$
where $\Delta_{\nu}$ is the excitation gap and
$v_{\nu}$ is the spin velocity in spin branch $\nu$.
However, for strong attractive interaction the low energy physics is
dominated by charge density fluctuations. 
This is because the spin wave fluctuations are fully suppressed by 
a large energy gap at low temperatures. 
This configuration is evidenced by the universality
class of finite-size correction to the ground state energy 
$E(L,N)-LE_0^{\infty}=-\frac{\pi\hbar C}{6L}\sum_{k=1}^4v_k$, where
the central charge $C=1$ for the $U(1)$ symmetry and charge velocities
$v_k\approx \frac{\hbar \pi n_k}{mk}(1+\frac{2}{|c|}A_k)$ with
$k=1,\ldots, 4$ are the charge velocities for unpaired fermions and
charge bound states.
For the singlet ground state, the spin velocity 
$v_{s} \approx \frac{\sqrt{5}|c|}{\sqrt{2}}(1+\frac{1}{3|\gamma|})$ in the spin-3/2
hyperfine branch, which is divergent due to the energy gap.

We find from the dressed energy equations (\ref{TBA-F}) that quantum phase
transitions driven by Zeeman splittings can be determined by three
independent external field-energy transfer relations 
\begin{eqnarray}
H_1 = \frac{5}{4}c^2+ u^{\rm 1}-u^{\rm 4}, \quad 
H_2=2c^2+ 2(u^{\rm 2}-u^{\rm 4}), \quad
H_3=\frac{7}{4}c^2+ 3(u^{\rm 3}-u^{\rm 4}).\label{mu-H}
\end{eqnarray}
The Fermi surfaces and charge bound states are fully controlled by the 
Zeeman splitting parameters.
The complete phase diagrams are determined by the
equations (\ref{mu-H}) and certain combinations of these equations.
Without loss of generality, we consider only terms up to order of
$1/|\gamma|$ in the following analysis.

\begin{figure}
{{\includegraphics [width=0.95\linewidth]{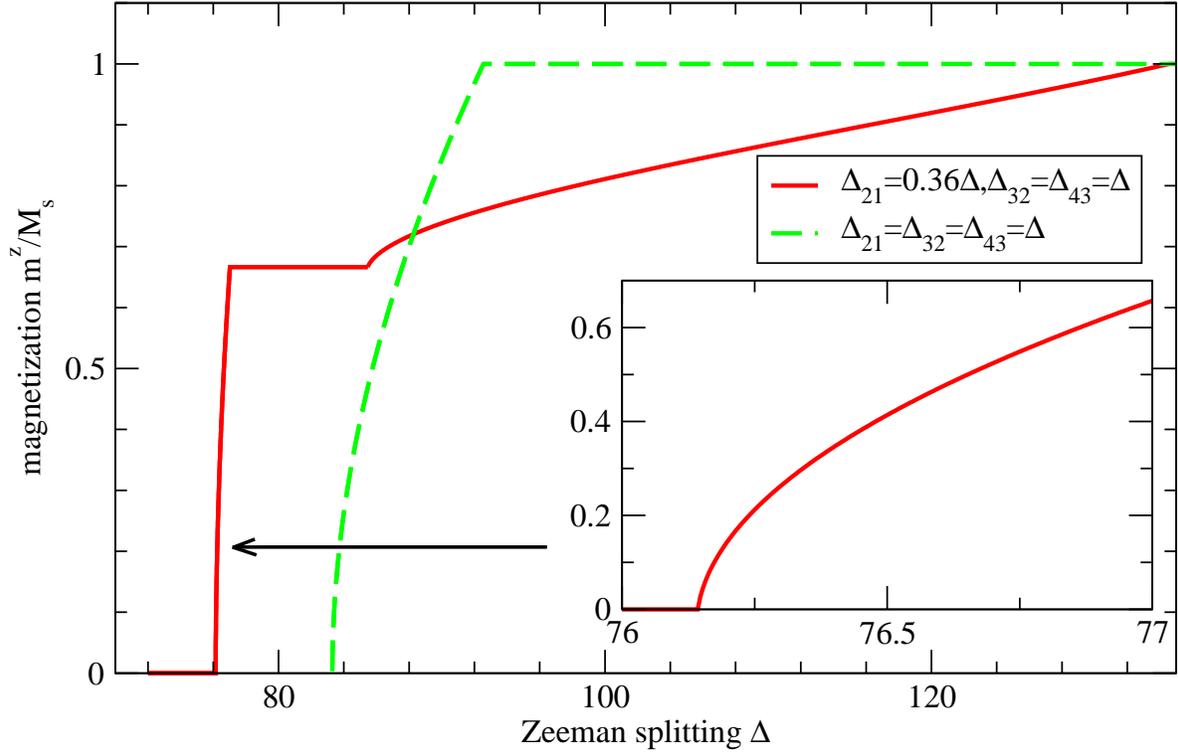}}}
\caption{Magnetization as a function of the Zeeman splitting parameter $\Delta$.
The curves shown, obtained from (\ref{mu-H}) with the effective
  chemical potentials given in (\ref{mu}), are for  linear Zeeman
  splitting (dashed line) with $c=-10$ and $n=1$ and unequally
  spaced Zeeman splitting (solid line) with $c=-8$ and $n=1$.
  The 2/3 magnetization
  plateau occurs through subtle tuning of the Zeeman splitting parameters. 
  Here, e.g., $\Delta_{21}=f\Delta_{43}=f\Delta_{32}$ with $f=0.36$. }
  \label{fig:mz-h}
\end{figure}

Using the energy transfer
relations (\ref{mu-H}), we find that linear Zeeman splitting
$\Delta$ lifts the $SU(4)$ symmetry to $U(1)^4$ symmetry for 
$|\gamma| \gg 1$, i.e., linear Zeeman splitting does not favor spin-neutral bound
states (recall part (A) in Fig~\ref{fig:states}). 
The lower critical
field $ \Delta_{c1}\approx
\frac{5c^2}{6}-\frac{n^2\pi^2}{384}(1+\frac{7}{18|\gamma|})$
diminishes the gap, thus the excitations becomes gapless. 
Using the definition of magnetization $m^z=M^z/M_s$ with
$M^z=\frac{3}{2}n_1+2n_2+\frac{3}{2}n_3$ and $M_s=\frac{3}{2}n$, we
see that in the vicinity of $\Delta_{c1}$, the system exhibits a
linear field-dependent magnetization of the form $m^z=192(\Delta-\Delta_{c1})/(n\pi^2)$ 
with a finite susceptibility
$\chi =\partial m^z/\partial \Delta \approx 192/(n\pi^2)$.
This result provides a testing ground for low energy field theory \cite{Tsvelik-book,Giamarchi-b}.  
When the Zeeman splitting $\Delta $ is greater than the upper critical field
$\Delta_{c2}\approx \frac{5c^2}{6}+\frac{2n^2\pi^2}{3}(1-\frac{2}{9|\gamma|})$ 
the system is fully-polarized into a normal Fermi liquid. For the intermediate
regime, $\Delta_{c1}< \Delta <\Delta_{c2}$, the quarteting state and
unpaired fermions coexist.  
The phase transition at the critical point
$\Delta_{c2}$ belongs the same linear field-dependent universality class.
A plot of the magnetization vs Zeeman splitting is given in Figure \ref{fig:mz-h}.

\begin{figure}
{{\includegraphics [width=0.95\linewidth]{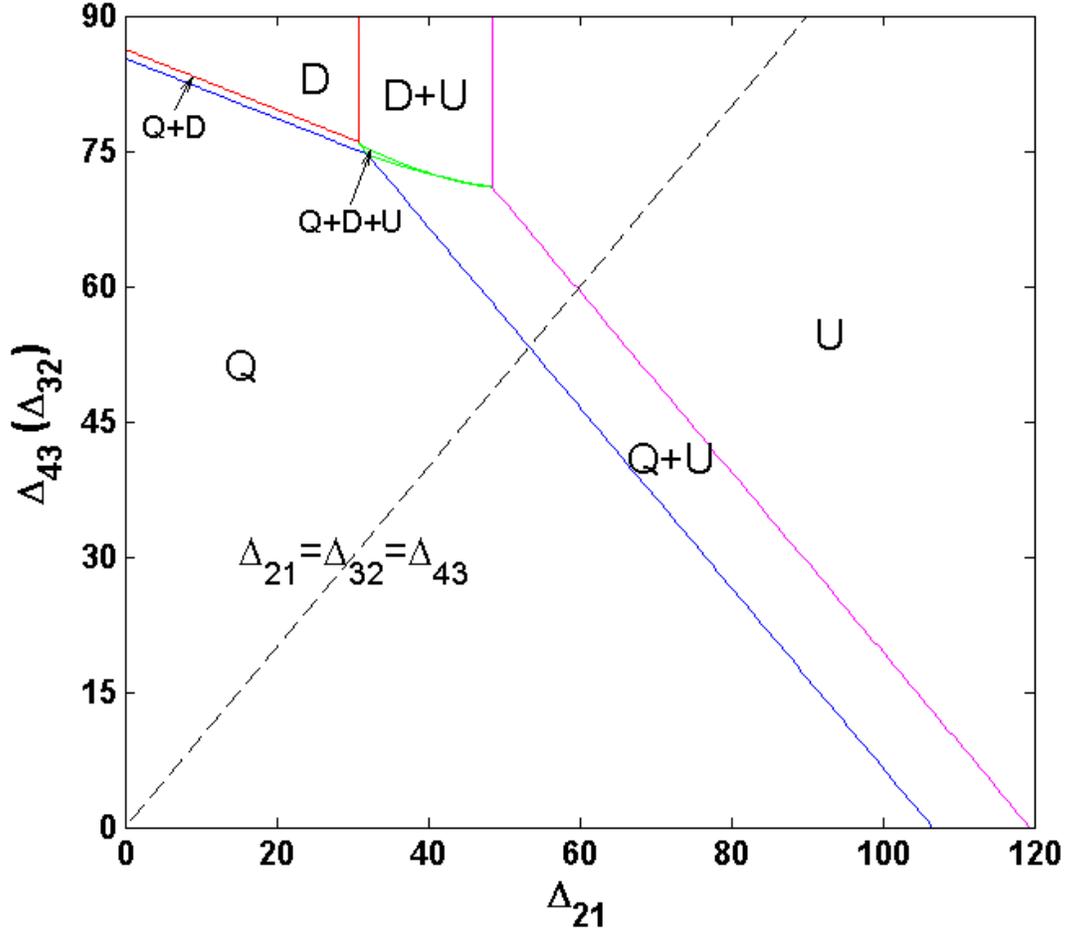}}}
\caption{Typical phase diagram in the $\Delta_{21}-\Delta_{43}(\Delta_{32})$ plane
  with $\Delta_{43}=\Delta_{32}$
  in the strong coupling regime $c=-8$ and $n=1$. 
  In this case we see clearly that Zeeman
  splitting can trigger a quantum phase transition from a quarteting 
  state (phase Q) into two spin-2 bound pairs (phase D), as depicted schematically in 
  part (C) of Figure \ref{fig:states}. Other phases involved are unpaired fermions (U) 
  and mixed phases.
  }\label{fig:mz-d21-d}
\end{figure}

For nonlinear Zeeman splitting the quarteting state can break into
two spin-2 bound pairs as depicted in part (C) of Figure \ref{fig:states}.  
In order to trigger such a paired phase, we let $\Delta_{43}=\Delta_{32}$
in the relation (\ref{H-Z}). 
In Figure \ref{fig:mz-d21-d} we demonstrate
the resulting interplay between the quantum phases of quarteting states (phase
$Q$), spin-2 pairs (phase $D$) and unpaired fermions (phase $U$). 
We see clearly that the quarteting states are stable for
$\Delta_{21}<\frac{5}{3}c^2-\frac{n^2\pi^2}{192}(1+\frac{7}{18|\gamma|})-\Delta_{43}$
and
$\Delta_{43}<\frac{4}{3}c^2-\frac{n^2\pi^2}{192}(1+\frac{13}{36|\gamma|})-\frac{\Delta_{21}}{3}$. 
The phase diagram shown in Figure \ref{fig:mz-d21-d} is determined by the
first two equations of (\ref{mu-H}) and 
$\Delta_{21}/2=c^2/4+\mu^{\rm 1}-\mu^{\rm 2}$ describing the mixed $D+U$ phase. 
In order to clearly see magnetization plateaux, we choose a
simple linear relation $\Delta_{21}=f\Delta_{43}$. 
We then find that
$f<\frac{3}{7}(1-\frac{5\pi^2}{14\gamma^2})$ and magnetization plateaux
$m^z=0,{2M_s}/{3},M_s$ occur, where $M_s=3n/2$ is the saturation
magnetization (see Figure \ref{fig:mz-h}). 
This is obvious because the lines fixed by the slope $f$ simultaneously pass
the $Q$, $D$ and $U$ phases.
Moreover, if the Zeeman splitting $\Delta_{32}$ is large enough,  
the phases $D$ and $D+U$ form the same phase diagram for spin-1/2
interacting fermions with polarization \cite{GBLB}.

\begin{figure}
{{\includegraphics [width=0.95\linewidth]{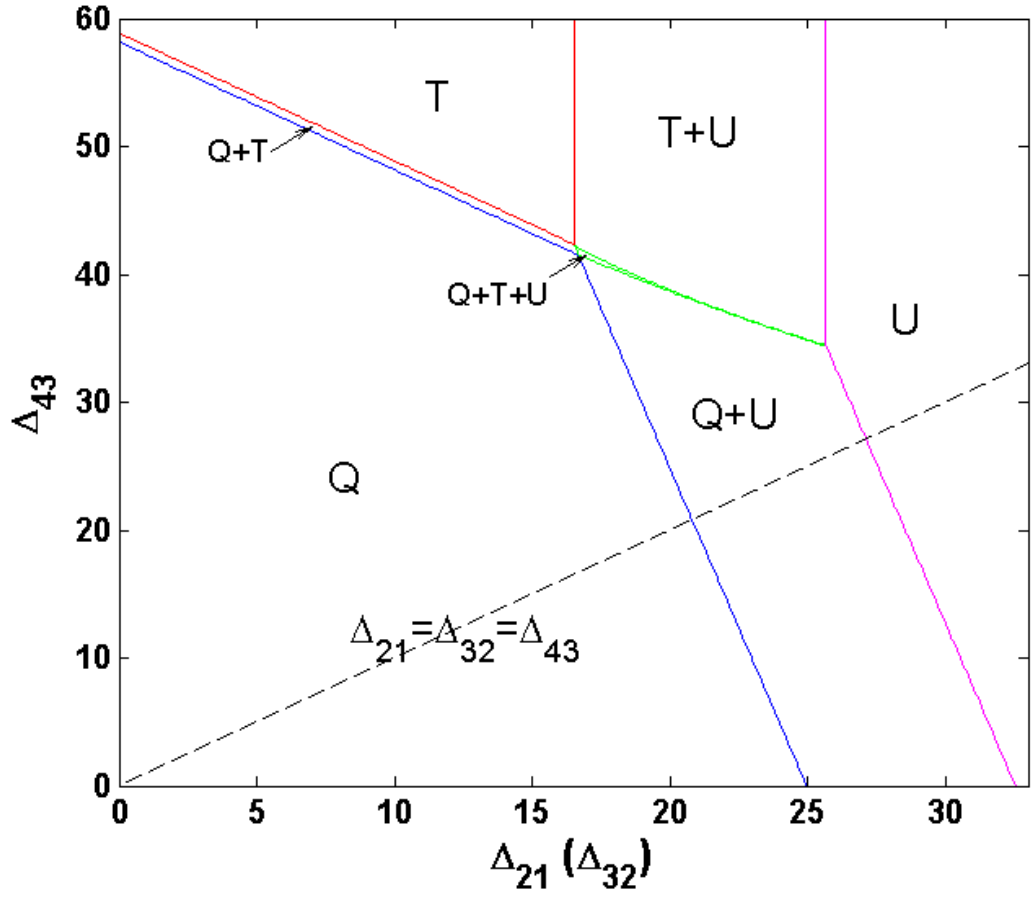}}}
\caption{Typical phase diagram in the $\Delta_{21}(\Delta_{32})-\Delta_{43}$ plane
  with $\Delta_{21}=\Delta_{32}$ in the strong coupling regime $c=-5$ and $n=1$. 
  In this case we see clearly that Zeeman
  splitting can trigger a quantum phase transition from quarteting
  states (phase Q) to trions (phase T), as depicted in part (B) of 
  Figure \ref{fig:states}. Other phases involved are unpaired fermions (U) 
  and mixed phases.}
  \label{fig:mz-d43-d}
\end{figure}

If we set $\Delta_{21}=\Delta_{32}$ in the relation
(\ref{H-Z}), the Zeeman parameters may trigger a phase transition from
quarteting states into trions (recall part (B) in Figure \ref{fig:states}). 
Figure \ref{fig:mz-d43-d} shows an exact
phase diagram in the $\Delta_{21}(\Delta_{32})-\Delta_{43}$ plane. 
Varying the Zeeman splitting $\Delta_{43}$ and $\Delta_{21}$ 
reveals smooth phase transitions from quarteting states into
trions (phase $T$) or a normal Fermi liquid (phase U). 
Similarly, we choose
$\Delta_{43}=f\Delta_{21}$ and see that magnetization plateaux
$m^z=0,{M_s}/{3},M_s$ occur if
$f>\frac{5}{2}(1+\frac{\pi^2}{18\gamma^2})$. 
The critical points for the plateaux can be analytically determined from the first
and third equations in (\ref{mu-H}) and an additional equation
$\Delta=\frac{2}{3}c^2+\mu^{\rm 1}-\mu^{\rm 3}$ for the mixed  $T+U$
phase. 
Actually, if $\Delta_{43}$ is large enough, the model reduces
to attractive fermions with three hyperfine levels \cite{GBLZ}. 
We point out that all phase transitions are of second order with a
linear field-dependent magnetization in the vicinities of critical points.

To conclude, we have presented unified exact results for complete
phase diagrams and quantum phase transitions from quarteting states
into spin-2 paired states and trions in 1D $SU(4)$ fermions with population imbalance.
The ground state properties and magnetic effects provide a testing ground for
low energy effective field theory and a bench mark for experiments 
with multicomponent ultracold fermionic atoms.

This work has been supported by the Australian Research Council. We
thank Y.-Q. Li and M. Oshikawa for discussions. C.L. thanks Yu.S. Kivshar
for support.

\clearpage

\end{document}